\documentclass[conference]{IEEEtran}
\IEEEoverridecommandlockouts

\usepackage{cite}
\usepackage{amsmath,amssymb,amsfonts}
\usepackage{algorithmic}
\usepackage{graphicx}
\usepackage{textcomp}
\usepackage{xcolor}
\usepackage{comment}
\usepackage{subcaption}
\usepackage{booktabs}
\usepackage{quantikz}
\usepackage{accessibility}
\usepackage{tikz}
\usepackage{hyperref}
\usepackage[capitalize]{cleveref}
\usetikzlibrary{
  positioning, arrows.meta, calc,
  shapes.geometric, fit, backgrounds,
  decorations.pathreplacing, calligraphy
}

\def\BibTeX{{\rm B\kern-.05em{\sc i\kern-.025em b}\kern-.08em
    T\kern-.1667em\lower.7ex\hbox{E}\kern-.125emX}}

\colorlet{Cconv}{blue!22}
\colorlet{Cres}{orange!30}
\colorlet{Cpool}{teal!25}
\colorlet{Cmlp}{violet!22}
\colorlet{Cout}{red!20}

\tikzset{
  base/.style={
    draw=black!60, rectangle, rounded corners=3pt,
    minimum width=10.5cm, minimum height=8.5mm,
    align=center, inner sep=3pt, font=\Large
  },
  conv/.style={base, fill=Cconv},
  resconv/.style={base, fill=Cres},
  pool/.style={base, fill=Cpool},
  mlp/.style={base, fill=Cmlp},
  outbox/.style={base, fill=Cout},
  smallout/.style={
    draw=black!60, rectangle, rounded corners=3pt,
    minimum width=3.0cm, minimum height=7.5mm,
    align=center, fill=Cout, font=\Large
  },
  arr/.style={-{Stealth[length=2.2mm,width=1.4mm]}, line width=0.75pt},
  resarr/.style={arr, dashed, gray!70},
  dimtag/.style={font=\footnotesize\itshape, text=gray!75},
  seclabel/.style={font=\scriptsize\bfseries, text=black!45},
}

\crefname{section}{Sec.}{Secs.}
\Crefname{section}{Sec.}{Secs.}
\crefname{subsection}{Sec.}{Secs.}
\Crefname{subsection}{Sec.}{Secs.}
\crefname{subsubsection}{Sec.}{Secs.}
\Crefname{subsubsection}{Sec.}{Secs.}
\Crefname{figure}{Fig.}{Figs.}
\Crefname{table}{Tab.}{Tabs.}
\crefname{definition}{Def.}{Defs.}
\Crefname{definition}{Def.}{Defs.}

\begin{document}

\title{Fidelity-Aware Scheduling of Quantum Circuits on Multi-QPU Systems}

\author{
\IEEEauthorblockN{Innocenzo Fulginiti\IEEEauthorrefmark{1}, Antonio Tudisco\IEEEauthorrefmark{2}, Salvatore Zammuto \IEEEauthorrefmark{1}, Patrick Hopf \IEEEauthorrefmark{1} \IEEEauthorrefmark{3}, Deborah Volpe \IEEEauthorrefmark{4}, \\ Helmut Seidl \IEEEauthorrefmark{1}, Giovanna Turvani \IEEEauthorrefmark{2}, Robert Wille \IEEEauthorrefmark{1} \IEEEauthorrefmark{3}, Christian B.~Mendl \IEEEauthorrefmark{1}, Martin Schulz \IEEEauthorrefmark{1}}

\vspace{0.5em}
\IEEEauthorblockA{\IEEEauthorrefmark{1}Technical University of Munich, Munich, Germany}
\IEEEauthorblockA{\IEEEauthorrefmark{2}Department of Electronics and Telecommunications, Politecnico di Torino, Turin, Italy}
\IEEEauthorblockA{\IEEEauthorrefmark{3}MQSC, Garching near Munich, Germany}
\IEEEauthorblockA{\IEEEauthorrefmark{4}Istituto Nazionale di Geofisica e Vulcanologia, Rome, Italy}

\vspace{0.5em}
\IEEEauthorblockA{
    \{antonio.tudisco, giovanna.turvani\}@polito.it;\\ 
    \{innocenzo.fulginiti, salvatore.zammuto, patrick.hopf, helmut.seidl, robert.wille, christian.mendl, martin.w.j.schulz\}@tum.de;\\ 
    deborah.volpe@ingv.it; }
}

\maketitle

\begin{abstract}

High Performance Computing-Quantum Computing (HPCQC) platforms expose multiple Quantum Processing Units (QPUs) that may differ in size, topology, native gates, and noise characteristics. For current noisy devices, errors compound along the compiled circuits quickly, and minimizing them, that is, maximizing the circuits' execution fidelity, is essential for reliable results.
Fidelity depends on the compilation to a specific target device: the same high-level circuit may produce different executables and, therefore, different expected fidelities across QPUs.
We present a low-overhead fidelity-aware scheduling framework for multi-QPU systems based on a Graph Neural Network (GNN) that estimates, before compilation, the expected fidelity of each circuit on each available QPU. Then, a tunable scheduler uses these estimates to control the trade-off between execution fidelity and parallelism. Results show that this framework allows for approximating an exhaustive fidelity-based assignment, saving computational resources compared to a brute-force approach that compiles each circuit on every device.

\end{abstract}



\section{Introduction}
\label{sec:introduction}
Quantum computers are becoming part of broader infrastructures, including cloud platforms and High-Performance Computing (HPC) environments~\cite{alexeev2021quantum,beck2024integrating}.
In these settings, users do not necessarily interact with a single isolated quantum processor. Instead, they submit tasks to a shared system that may expose multiple quantum processing units (QPUs). These QPUs may differ in technology, size, connectivity, native gate set, and noise characteristics~\cite{alexeev2021quantum,zhu2025quantum}.
This opens opportunities for parallel execution, but also introduces new challenges in workload placement and scheduling \cite{ravi2021adaptive, wang2024qoncord, nguyen2025qfor, giortamis2025qos, raj2026quantumintegratedhighperformancecomputing}.
A natural workload model in such environments is a task composed of multiple circuits. Such a task may contain circuits with different widths, depths, gate compositions, and multi-qubit interaction patterns. This structural diversity makes device selection a circuit-dependent decision: different circuits may be best executed on different QPUs. 
Indeed, even when different QPUs provide a comparable number of qubits, they may differ in their physical topology, supported native gates, gate error rates, and coherence properties~\cite{zhu2025quantum}.
These hardware characteristics directly affect the resulting compiled circuit.
A high-level circuit can lead to substantially different compiled implementations, and, as a consequence, different expected execution fidelities, depending on the selected target device \cite{qiskit2019, molavri2022, cheng2024}.
This issue is particularly evident in the Noisy Intermediate-Scale Quantum (NISQ) regime, where current quantum devices are affected by gate noise and limited coherence~\cite{preskill2018nisq}. Therefore, the quality of a circuit execution strongly depends on the accumulated error of the compiled circuit. For this reason, fidelity is a central metric when deciding where a circuit should be executed, particularly in a heterogeneous multi-QPU system, where selecting an unsuitable device can degrade the quality of the result \cite{Hopf}.
Fidelity can only be estimated once the circuit has been compiled for a specific device, as it is influenced by the actual gate decomposition and qubit mapping imposed by that hardware~\cite{qiskit2019,zhu2025quantum}.
An exhaustive multi-device compilation strategy, in which each circuit is compiled for every available QPU to estimate the expected fidelity using hardware calibration data, introduces substantial overhead. Indeed, circuit compilation is a process that tends to be expensive in terms of runtime costs~\cite{zhu2025quantum}. Moreover, only one compiled version will ultimately be executed, while all the other versions are generated only to support the assignment decision. On the other hand, optimizing fidelity alone is not always sufficient in a multi-QPU system, as a purely fidelity-driven policy may concentrate circuits on a small subset of devices, leaving other QPUs underutilized.

In this paper, we propose a framework that avoids exhaustive multi-device compilation by determining the most suitable QPU for each circuit before device-specific compilation.
The core idea is to exploit a Graph Neural Network (GNN)-based model that, given a high-level quantum circuit, predicts the expected execution fidelity of the post-compilation circuit on each available QPU. These predictions are then used by a tunable scheduler, which assigns circuits to QPUs according to a user-defined setting that controls how strongly the system should prioritize fidelity over parallelism. Once a target device has been selected, the circuit is compiled only for that device, thereby reducing the compilation overhead associated with exhaustive device selection.
We evaluate the proposed framework on an emulated multi-QPU environment derived from real IQM superconducting devices. The results show that the GNN predictor estimates post-compilation circuit fidelities with low error, and that the proposed scheduler closely approaches the fidelity of ground-truth-based assignment while avoiding exhaustive compilation across all devices. Moreover, by tuning the fidelity weight, the framework can move from balanced device utilization to fidelity-oriented assignments, exposing a practical trade-off between parallelism and execution quality.

The rest of the paper is organized as follows.
\cref{sec:motivation} motivates the need for fidelity-aware and tunable scheduling in multi-QPU systems.
\cref{sec:related_works} reviews the relevant literature.
\cref{sec:methodology} presents the proposed framework, including the fidelity prediction model and the tunable scheduling policy.
\cref{sec:settings} describes the emulated multi-QPU environment, the dataset, and the training procedure adopted for the evaluation.
\cref{sec:results} reports the obtained results and compares the proposed approach against baseline assignment strategies.
Finally, \cref{sec:conclusions} concludes the paper.


\section{Motivation}
\label{sec:motivation}
Consider a heterogeneous High Performance Computing-Quantum Computing (HPCQC) system exposing a set of QPUs with different hardware characteristics. Users submit tasks that consist of batches of \(k\) circuits. The goal of the system is to assign these circuits to the available QPUs, preserving high execution fidelity while also exploiting the degree of parallelism requested by the user.

\subsection{Circuit-Dependent Fidelity}
In the NISQ setting, the execution quality of a circuit $c$ on a device $d$ can be quantified in terms of the \emph{expected fidelity} $f_{c,d}$, which can be estimated by accumulating the success probabilities of all operations in the circuit as

\begin{equation}
\label{eq:fidelity}
    f_{c,d}
        = \prod_{g \in G_{c,d}}
          \bigl(1 - \varepsilon_d(g, Q_g)\bigr)
\end{equation}

where $G_{c,d}$ is the sequence of operations of the circuit $c$ compiled for $d$, $Q_g$ is the set of qubits on which the gate $g \in G_{c,d}$ acts, and $\varepsilon_d(g, Q_g)$ denotes the error rate of gate $g$ acting on $Q_g$ for device $d$, i.e.\ the gate error rate for gates and the readout error rate for measurements.
\cref{eq:fidelity} accumulates the calibrated success probabilities of gate and readout operations and, like the Estimated Success Probability (ESP) \cite{Hopf} commonly used for device selection, it neglects idle-time decoherence over the schedule duration, crosstalk between concurrently driven qubits, calibration drift between snapshots, and shot noise. It is therefore an approximation of the fidelity that would be measured on hardware, and is adopted here because it is the quantity that exhaustive multi-device compilation itself computes: the comparison between our predictor and the exhaustive baseline is unaffected by this choice, since both are evaluated against the same target. Absolute fidelity values should nonetheless be read as relative indicators of device suitability rather than as predictions of measured outcomes.

The expected fidelity of a circuit cannot be estimated from the high-level circuit alone, since it depends on both the device-specific compiled implementation and the hardware characteristics of the target QPU, such as per-qubit and per-gate error rates. During the compilation process, the logical circuit must be mapped and routed onto the physical qubits of the selected QPU. Furthermore, quantum gates need to be translated and decomposed into the native gate set supported by that device. As a result, the same high-level circuit can lead to different compiled circuits on different QPUs. A device whose topology matches the circuit connectivity may require fewer routing operations, while another device may introduce additional gates due to limited connectivity or a different native gate set. These device-specific transformations affect gate count and, consequently, accumulated error, which determines the expected execution fidelity.
Moreover, even two QPUs sharing the same topology and native gate set can yield substantially different execution fidelities for the same circuit if their noise profiles differ.

\subsection{Cost of Exhaustive Device Selection}
A straightforward method for fidelity-aware assignment is to compile each circuit on each available QPU, evaluate the fidelity of each compiled version, and select the device with the highest fidelity. This strategy provides a strong hardware-aware baseline, since the decision is based on device-specific compiled circuits.
However, this approach incurs significant overhead, as quantum circuit compilation is computationally expensive and scales poorly with circuit size, with core subproblems such as qubit mapping and routing known to be NP-hard and often dominating compilation time in practice \cite{molavri2022, cheng2024, zhu2025quantum}. Moreover, for a batch of $k$ circuits and $D$ available QPUs, it requires $k \times D$ compilations, although only $k$ compiled circuits are eventually executed.
The remaining compiled versions are generated just for the assignment decision. This overhead grows with the number of circuits and the number of available devices. Therefore, reducing this overhead by predicting the expected execution fidelity of each circuit on each QPU before device-specific compilation is particularly beneficial. In this way, the circuit is compiled only for the selected QPU, preserving the idea of fidelity-aware assignment while avoiding compilation attempts for devices that are unlikely to be selected.

\subsection{Limits of Fidelity-First Scheduling and Parallelism Trade-offs}
The selection of the QPU with the highest predicted fidelity maximizes execution quality at the circuit level, but may not fully exploit a multi-QPU system. If many circuits in a batch are predicted to run best on the same device, a strict fidelity-first policy assigns most of the workload to that QPU, while other devices remain underutilized.
The opposite extreme is also undesirable. A policy that only balances the number of circuits across QPUs, such as \emph{round robin}, can improve device utilization but ignores device-dependent fidelity. As a result, it may assign circuits to QPUs on which they are expected to execute with substantially lower fidelity. Therefore, fidelity and parallelism represent competing objectives: improving one may reduce the other. 

To address this trade-off, our framework exposes a tunable scheduling parameter, i.e., the fidelity weight \(w \in [0,1]\). Lower values of \(w\) give more importance to parallel execution, while higher values prioritize the predicted fidelity. In the extreme case \(w = 0\), the scheduler optimizes workload distribution across the available QPUs, behaving similarly to a round-robin policy. In the extreme case \(w = 1\), each circuit is assigned to the QPU predicted to provide the highest fidelity. Intermediate values allow the scheduler to trade limited fidelity loss for better distribution across devices.



\section{Related Works}
\label{sec:related_works}
Scheduling quantum circuits across the QPUs of an HPCQC system based on fidelity estimates is a problem that touches several research areas, including device selection, fidelity estimation, and resource scheduling. However, to the best of our knowledge, none of the existing works provides a complete workflow for assigning a batch of circuits to the available QPUs based on pre-compilation fidelity estimates, while avoiding the overhead of multi-device compilation.

A closely related direction is automatic quantum device selection before compilation. In \cite{salm2023select, quetschlich2025mqtpredictor, tudisco2025gnn}, it is studied how quantum computers can be selected from the input circuit before executing a full compilation workflow to maximize various metrics, including fidelity. These works share our goal of avoiding expensive brute-force compilation across multiple devices. However, they mainly focus on selecting a target device for an individual circuit, treating the problem as a classification task that returns the device with the highest fidelity.
In contrast, our framework proposes a scheduler that exploits a Machine Learning (ML) model to estimate the expected fidelity of each circuit on every available QPU, dispatching circuits according to both fidelity information and QPU workload.

A previous work, \cite{11662286}, presents a GNN-based model that predicts the ESP of a quantum circuit for multiple QPUs. It has been shown that a graph representation of the \emph{uncompiled} circuit yields more accurate estimates than feature-based regressors. However, that work is confined to the \emph{estimation} task, missing all aspects related to dispatching the circuit across different QPUs. In addition, the model employed here additionally refines the graph encoder with a richer readout, in which the mean, sum, and maximum of the node embeddings are concatenated rather than reduced through a single pooling function.


ML has also been explored for predicting the reliability of quantum circuits. The work in \cite{wang2022quest} uses a graph transformer to estimate circuit fidelity from graph representations of quantum circuits. On a different line, \cite{mao2025qfid} uses LSTM networks to model circuit fidelity from tokenized gate sequences. These approaches show that ML models can approximate expensive fidelity estimation procedures and capture non-trivial properties of noisy circuit execution. Our work builds on this general idea, but uses fidelity prediction as part of a resource management pipeline to drive scheduling decisions across heterogeneous QPUs.

Scheduling and resource management have received increasing attention as quantum hardware becomes accessible through shared infrastructures.
In \cite{ravi2021adaptive}, resource management for quantum clouds is proposed, considering features like queue times, fidelity trends, and calibration constraints. In \cite{giortamis2025qos}, the authors present a tool that manages quantum resources through hardware-agnostic execution, error mitigation, multiprogramming, and scheduling. The work in \cite{nguyen2025qfor} formulates fidelity-aware orchestration across heterogeneous quantum nodes using deep reinforcement learning, balancing execution fidelity, given an already compiled circuit, and execution time. The authors of \cite{wang2024qoncord} address multi-device scheduling for variational quantum algorithms by distributing different optimization phases across devices to reduce queueing
delays and improve resource usage. These systems address important scheduling and orchestration challenges in quantum HPCQC systems.
In contrast, our framework performs device selection before device-specific compilation, predicting post-compilation fidelity directly from the high-level circuit representation rather than from compiled instances across all candidate QPUs.

Other works exploit parallelism by executing multiple circuits concurrently on a single quantum processor, partitioning physical qubits so that multiple programs can run simultaneously while accounting for fidelity and resource utilization \cite{liu2021qucloud, niu2023qumc}.

Finally, \cite{tejedor2025qdislib} addresses the execution of large quantum circuits on resource-constrained devices by exploiting distributed quantum circuit cutting for hybrid quantum-classical HPC systems. Therefore, they decompose large circuits into smaller subcircuits that can be distributed and executed across heterogeneous resources.

Overall, our work differs from prior approaches by combining pre-compilation fidelity prediction with tunable multi-circuit scheduling in a heterogeneous multi-QPU setting. The main contributions of this paper are:
    

\begin{itemize}
    \item an end-to-end framework that integrates pre-compilation fidelity estimation with multi-circuit allocation;

    \item a tunable scheduling policy that assigns batches of independent circuits to heterogeneous QPUs through a single parameter $w \in [0,1]$, having round-robin and fidelity maximization as its two extremes;


    \item an experimental evaluation on an emulated multi-QPU environment derived from real IQM superconducting devices.
\end{itemize}


\section{Methodology}
\label{sec:methodology}
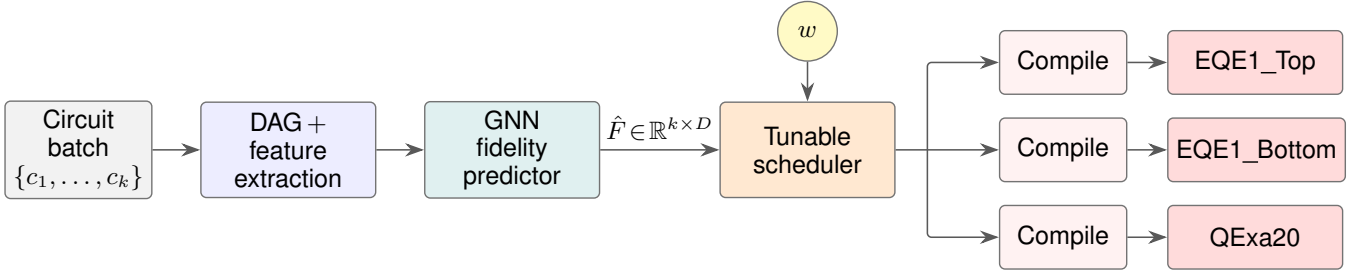
\begin{figure*}[t]
    \centering
    \resizebox{\linewidth}{!}{
\begin{tikzpicture}[
    font=\sffamily\small,
    >={Stealth[length=2.4mm, width=1.7mm]},
    node distance=4mm and 6mm,
    every node/.style={align=center, inner sep=3pt},
    block/.style={
        rectangle, rounded corners=2pt,
        draw=black!55, line width=0.5pt,
        minimum height=12mm,
    },
    io/.style      ={block, fill=gray!10,   minimum width=18mm},
    pre/.style     ={block, fill=blue!7,    minimum width=22mm},
    model/.style   ={block, fill=teal!12,   minimum width=22mm},
    sched/.style   ={block, fill=orange!18, minimum width=22mm},
    compile/.style ={block, fill=red!5,   minimum height=8mm, minimum width=16mm},
    qpu/.style     ={block, fill=red!14,  minimum height=8mm, minimum width=22mm},
    knob/.style    ={
        circle, draw=black!55, line width=0.5pt,
        fill=yellow!28, minimum size=7.5mm, inner sep=0pt,
    },
    flow/.style    ={->, draw=black!65, line width=0.55pt},
    branch/.style  ={    draw=black!65, line width=0.55pt, rounded corners=1.5pt},
    arrlbl/.style  ={font=\scriptsize, color=black!75, fill=white, inner sep=1.2pt},
]

\node[io]                          (batch) {Circuit\\batch\\$\{c_1,\dots,c_k\}$};
\node[pre,    right=of batch]      (dag)   {DAG\,$+$\\feature\\extraction};
\node[model,  right=of dag]        (gnn)   {GNN\\fidelity\\predictor};
\node[sched,  right=15mm of gnn]   (sch)   {Tunable\\scheduler};
\node[knob,   above=5mm of sch]    (alpha) {$w$};

\node[compile, right=13mm of sch, yshift= 11mm] (c1) {Compile};
\node[compile, right=13mm of sch, yshift= 0mm]  (c2) {Compile};
\node[compile, right=13mm of sch, yshift=-11mm] (c3) {Compile};

\node[qpu, right=5mm of c1] (q1) {EQE1\_Top};
\node[qpu, right=5mm of c2] (q2) {EQE1\_Bottom};
\node[qpu, right=5mm of c3] (q3) {QExa20};

\draw[flow] (batch) -- (dag);
\draw[flow] (dag)   -- (gnn);
\draw[flow] (gnn)   -- (sch)
    node[midway, above]{$\hat{F}\!\in\!\mathbb{R}^{k\times D}$};
\draw[flow] (alpha) -- (sch);

\coordinate (fork) at ($(sch.east)+(4mm,0)$);
\draw[branch] (sch.east) -- (fork);
\draw[flow,   rounded corners=1.5pt] (fork) |- (c1.west);
\draw[flow]                          (fork) -- (c2.west);
\draw[flow,   rounded corners=1.5pt] (fork) |- (c3.west);

\draw[flow] (c1) -- (q1);
\draw[flow] (c2) -- (q2);
\draw[flow] (c3) -- (q3);

\end{tikzpicture}
    }
    \caption{End-to-end pipeline of the proposed framework. A user-submitted batch of $k$ high-level quantum circuits is converted into Directed Acyclic Graphs with per-node feature vectors and fed to the GNN fidelity predictor, which produces an estimated fidelity matrix $\hat{F}$ over all $k \times D$ circuit-device pairs. The tunable scheduler, parameterized by the fidelity weight $w$, routes each circuit to one of the available QPU backends; each assigned circuit is then transpiled against the target device's native gate set and connectivity, and then executed on the corresponding backend.}
    \label{fig:pipeline}
\end{figure*}

As previously mentioned, this work aims to maximize the overall fidelity of a batch of circuits while improving resource utilization. To this end, we develop a framework comprising three main steps: a pre-processing step that translates the input circuits into a format suitable for our model (\cref{sec:meth-preproc}); an ML model, specifically a GNN, that estimates the fidelity of the circuits on the available QPUs prior to compilation (\cref{sec:meth-model}); a scheduler that, based on the model's predictions, assigns the circuits in the batch to the available QPUs according to the selected policy (\cref{sec:meth-scheduler}). An overview of the end-to-end pipeline is shown in Fig.~\ref{fig:pipeline}.

\subsection{Pre-processing}
\label{sec:meth-preproc}
\begin{figure*}
    \centering
    \includegraphics[width=0.8\linewidth]{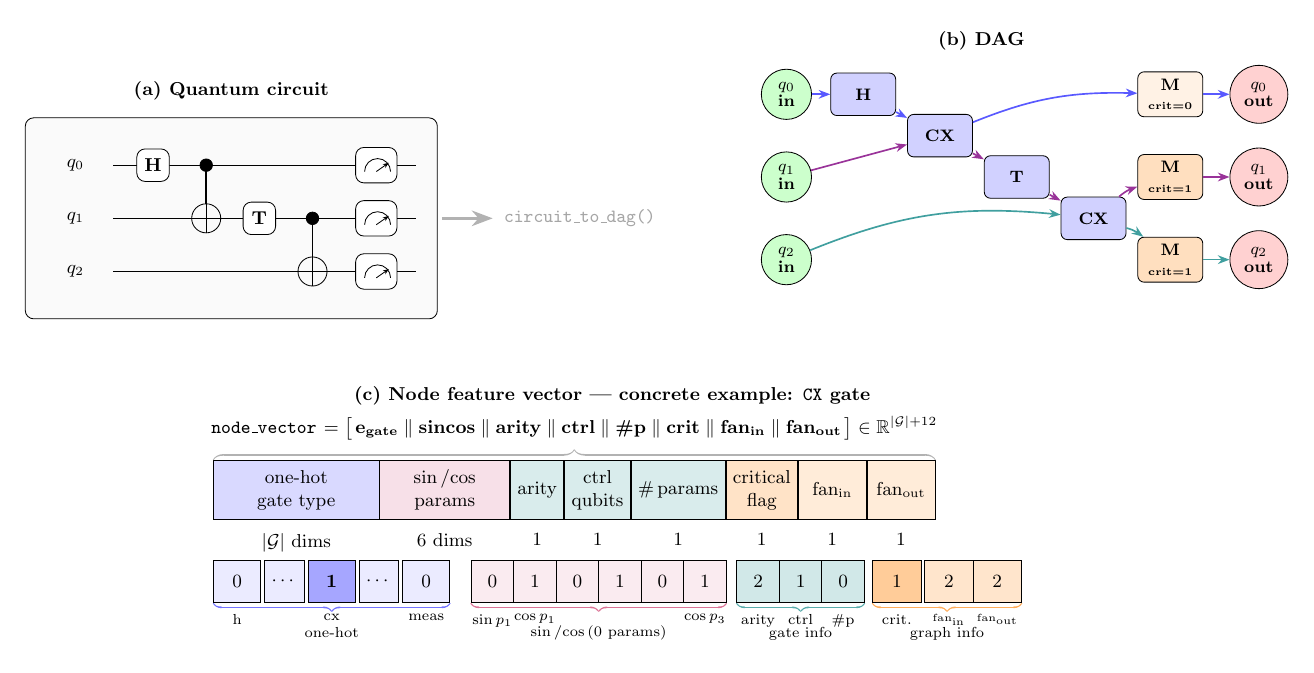}
    \caption{Conversion from circuit to DAG. A quantum circuit (a) is translated into a DAG (b), whose nodes are gates and measurements and whose edges encode data dependencies; each node is then mapped to the feature vector of (c), shown for a \texttt{CX} gate.}
    \label{fig:circuit-dag}
\end{figure*}

Before a quantum circuit can be processed by a GNN model, it must pass through a pre-processing step in which it is converted into a graph representation with numerical attributes, as shown in Fig.~\ref{fig:circuit-dag}. Quantum circuits can be naturally translated into a graph formulation as Directed Acyclic Graphs (DAGs), where each node represents a quantum gate or operation and each directed edge encodes a data dependency between operations, reflecting the order in which they must be executed. This representation preserves the structure of the circuit in a form that can be directly exploited by our GNN model.
Once the DAG is constructed, each node is mapped to a fixed-length numerical feature vector of 40 elements. Specifically, the first 28 elements encode the quantum operation type in one-hot format, following the QASM3~\cite{qasm} gate set. The subsequent 6 elements represent the sine and cosine of the rotation angle parameters associated with the gate, to allow encoding the angle parameters and their periodicity into a $[-1, 1]$ interval. The remaining 6 elements capture structural properties of the node, namely the number of qubits involved in the operation, the number of control qubits, the number of angle parameters, a binary flag indicating whether the gate lies on the device-agnostic critical path (i.e., the longest path in the DAG), the number of predecessors, and the number of successors.

\subsection{Model Definition}
\label{sec:meth-model}
Once the circuit has been translated as a numerical DAG, it is fed to a GNN model. This model consists of a graph encoder, which processes the input graph by aggregating information from neighboring nodes, and an MLP, which predicts the fidelity of the circuit on the target devices.

The graph encoder consists of a set of SAGE convolutional layers \cite{hamilton2018inductiverepresentationlearninglarge} organized into two distinct stacks: an initial series of SAGE convolutional layers applied directly to the input graph, and a subsequent set arranged in a residual configuration. SAGE convolutional layers are adopted to enable inductive learning, allowing the model to generalize to unseen quantum circuits.
Each convolutional layer is then followed by a normalization and regularization block, consisting of a Graph Normalization layer, a Leaky ReLU activation function, and a dropout layer. This sequence has a triple effect: stabilizes training, introduces non-linearity, and mitigates overfitting.

To fully exploit the directional structure of the DAG, an optional technique known as \emph{bidirectional message passing} can be employed, in which each convolutional layer is applied in parallel to both the original graph and its reversed counterpart. When enabled, the forward direction captures the natural flow of quantum operations, from inputs to outputs, while the reverse direction allows each node to aggregate information about the downstream consequences of its operation, i.e., how its output influences the gates that follow. Whether to enable this bidirectional aggregation is a hyperparameter of the model.

After all convolutional layers, the resulting node embeddings are aggregated into a single graph-level representation through a \emph{mean-max-sum pooling} readout. This operator concatenates three distinct aggregations of the node embeddings: their element-wise mean, their element-wise maximum, and their element-wise sum. This design aims to exploit all three advantages of the pooling techniques considered: the mean captures the average structural pattern across the graph, the maximum preserves the most prominent features, and the sum provides a size-aware aggregate that retains information about the overall scale of the circuit.

The optimal model was identified by tuning the hyperparameters outlined in Table~\ref{tab:hyperparam}.

\begin{table}[t]
  \centering
  \caption{Hyperparameter Search Space.}
  \label{tab:hyperparam}
  \renewcommand{\arraystretch}{1.2}
  \resizebox{\columnwidth}{!}{
  \begin{tabular}{lll}
    \toprule
    \textbf{Hyperparameter} & \textbf{Range} & \textbf{Best} \\
    \midrule
    \texttt{hidden\_dim}            & $[32,\,256]$                        & $142$\\
    \texttt{num\_conv\_wo\_resnet}  & $[1,\,3]$                           & $3$\\
    \texttt{num\_resnet\_layers}    & $[1,\,9]$                           & $1$\\
    \texttt{dropout}                & $\{0.0,\,0.1,\,0.2\}$               & $0.0$\\
    \texttt{bidirectional}          & $\{\text{True},\,\text{False}\}$    & $\text{True}$\\
    \texttt{mlp\_units} &
    $\begin{aligned}
    &(), (32), (64), (128), (256), (512), (64,32), (128,32),\\
    &(128,64), (256,32), (256,64), (256,128), (512,256),\\
    &(512,128), (512,64), (512,32), (128,64,32),\\
    &(256,128,64), (512,256,128)
    \end{aligned}$ & $(512,256,128)$ \\
    \texttt{lr}                     & $\{10^{-4},\,10^{-3},\,10^{-2}\}$   & $10^{-3}$\\
    \bottomrule
  \end{tabular}
  }
\end{table}


\subsection{Scheduling}
\label{sec:meth-scheduler}

We now present our fidelity-aware scheduler that assigns circuits in the incoming batch to one of the available QPU backends.
The scheduler exploits the per-device fidelity estimates produced by the GNN to make informed dispatch decisions while simultaneously controlling the distribution of workload across devices.


\subsubsection{Fidelity-aware device selection}

We formulate device selection as a sequential assignment over a unified circuit queue shared by all backends.

Let $\mathcal{D} = \{d_1, \ldots, d_D\}$ denote the set of available devices, $\hat{f}_{c,d} \in [0,1]$ the GNN-predicted fidelity of circuit $c$ on device $d$, $\ell_d$ the number of circuits already assigned to $d$ in the current dispatch round (load), and $L = \sum_{d'} \ell_{d'}$ the total number of circuits dispatched so far. Devices compete on their current \emph{load share}

\begin{equation}\label{eq:load_share}
    s_d =
    \begin{cases}
        \ell_d / L, & L > 0,\\[2pt]
        0,          & L = 0 .
    \end{cases}
\end{equation}

Circuit $c$ is assigned to the device maximizing a composite score:

\begin{equation}\label{eq:score}
    d^*(c) = \arg\max_{d \in \mathcal{D}} \; w \cdot \hat{f}_{c,d} \;-\; (1-w) \cdot s_d
\end{equation}

where $w \in [0, 1]$ is the \emph{fidelity weight}.
Predicted fidelities already lie in $[0,1]$ and load shares are fractions, so both terms share a common scale without further normalization. Retaining the \emph{raw} fidelities is deliberate: it preserves the magnitude of each circuit's device preference, so that a device holding a $0.001$ fidelity edge barely outbids the load term whereas a $0.3$ edge dominates it, and circuits whose preferences differ in strength switch device at correspondingly different weights. At $w = 0$ the score reduces to $-s_d$, selecting the least-loaded device and recovering a uniform round-robin distribution; at $w = 1$ it reduces to $\hat{f}_{c,d}$, selecting the device with the highest predicted fidelity.

\paragraph*{Interpretation of $w$} Because the penalty grows with the actual share imbalance rather than with a normalized rank, dispatching a further circuit to an already-loaded device becomes progressively more expensive. Devices therefore fill until the marginal share gap offsets the fidelity gap: at equilibrium, for any two devices $a, b$,

\begin{equation}\label{eq:equilibrium}
    s_a - s_b \;\approx\; \frac{w}{1-w}\,\bigl(\bar{f}_{a} - \bar{f}_{b}\bigr), \; \qquad \bar{f_d} = \frac{1}{k}\sum_{c=1}^k \hat{f}_{c,d}.
\end{equation}
where $\bar{f_d}$ is the mean fidelity of the batch of circuits assigned to the device $d$.
The ratio $w/(1-w)$ thus acts as an explicit exchange rate between fidelity advantage and load share: a device may accumulate $w/(1-w)$ units of additional load share per unit of fidelity advantage. Consequently, circuits do not all change device at the same value of $w$. Each circuit switches at the weight at which its own fidelity advantage outweighs the load penalty, so the allocation changes gradually over the entire range $w \in (0,1)$, as confirmed experimentally in \cref{sec:meth-scheduler-results}.
This change is not equally fast at all weights. The exchange rate $w/(1-w)$ grows slowly for small $w$ and diverges as $w$ approaches $1$: circuits with a pronounced device preference are therefore reassigned already at low weights, which is where most of the attainable fidelity gain is obtained, while the remaining load balance is given up only close to $w = 1$. At $w = 1$ the load term disappears entirely and the assignment reduces to a pure fidelity argmax. This behavior at the endpoint follows directly from ignoring the load and is not an artifact of the scoring function.


\subsubsection{Dispatcher integration}

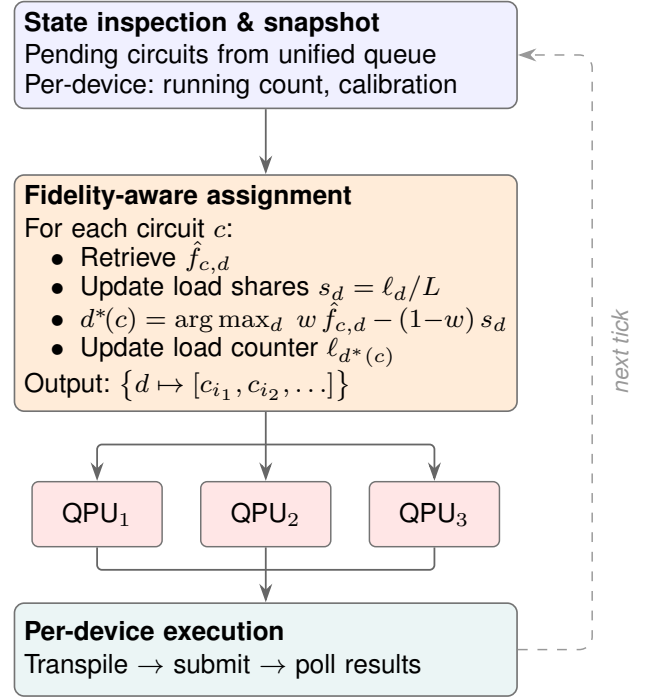
\begin{figure}[t]
    \centering
    \resizebox{0.95\columnwidth}{!}{%
\begin{tikzpicture}[
    font=\sffamily\footnotesize,
    >={Stealth[length=2mm, width=1.4mm]},
    node distance=7mm,
    every node/.style={align=center, inner sep=3pt},
    phase/.style={
        rectangle, rounded corners=3pt,
        draw=black!55, line width=0.5pt,
        text width=52mm, minimum height=10mm,
    },
    devbox/.style={
        rectangle, rounded corners=2pt,
        draw=black!55, line width=0.45pt,
        minimum width=14mm, minimum height=7mm,
        fill=red!10,
        font=\sffamily\footnotesize,
    },
    flow/.style={->, draw=black!60, line width=0.55pt},
]

\node[phase, fill=blue!6, align=left] (p1) {
    \textbf{State inspection \& snapshot}\\[1pt]
    Pending circuits from unified queue\\
    Per-device: running count, calibration
};

\node[phase, fill=orange!15, below=of p1, align=left] (p2) {
    \textbf{Fidelity-aware assignment}\\[1pt]
    For each circuit $c$:\\
    \quad$\bullet$\; Retrieve $\hat{f}_{c,d}$ \\
    \quad$\bullet$\; Update load shares $s_d = \ell_d / L$ \\
    \quad$\bullet$\; $d^{*}\!(c) = \arg\max_d\;
        w\,\hat{f}_{c,d} - (1\!-\!w)\,s_d$ \\
    \quad$\bullet$\; Update load counter $\ell_{d^*(c)}$\\[1pt]
    Output: $\bigl\{d \mapsto [c_{i_1}, c_{i_2}, \ldots]\bigr\}$
};

\node[devbox] (d2) at ($(p2.south)+(0,-11mm)$) {QPU$_2$};
\node[devbox, left=4mm of d2] (d1) {QPU$_1$};
\node[devbox, right=4mm of d2] (d3) {QPU$_3$};

\node[phase, fill=teal!8, align=left] (p3) at ($(d2.south)+(0,-11mm)$) {
    \textbf{Per-device execution}\\[1pt]
    Transpile $\to$ submit $\to$ poll results
};


\draw[flow] (p1) -- (p2);

\coordinate (fork) at ($(p2.south)+(0,-3.5mm)$);
\draw[draw=black!60, line width=0.55pt] (p2.south) -- (fork);
\draw[flow, rounded corners=1.5pt] (fork) -| (d1.north);
\draw[flow]                         (fork) --  (d2.north);
\draw[flow, rounded corners=1.5pt] (fork) -| (d3.north);

\coordinate (join) at ($(p3.north)+(0,3.5mm)$);
\draw[draw=black!60, line width=0.55pt, rounded corners=1.5pt]
    (d1.south) |- (join);
\draw[draw=black!60, line width=0.55pt]
    (d2.south) -- (join);
\draw[draw=black!60, line width=0.55pt, rounded corners=1.5pt]
    (d3.south) |- (join);
\draw[flow] (join) -- (p3.north);

\draw[->, draw=black!40, line width=0.45pt, dashed, rounded corners=4pt]
    (p3.east) -- ++(8mm,0) |- (p1.east);

\node[font=\sffamily\scriptsize\itshape, text=black!40, rotate=90]
    at ($(p3.east)!0.5!(p1.east) + (11mm,0)$) {next tick};

\end{tikzpicture}%
    }
    \caption{Logic of a single dispatch tick within the QMS. The broker constructs an immutable snapshot of the queue and per-device state, invokes the scheduling policy to produce a device-to-circuit mapping, and dispatches each circuit to its assigned backend for transpilation and execution. The dashed arrow indicates the tick-based cycle.}
    \label{fig:dispatch-logic}
\end{figure}

The scheduling policy described above is embedded in a Quantum Meta-Scheduler (QMS) that orchestrates the full circuit lifecycle within an HPC allocation.
The QMS operates a continuous dispatch loop over a unified circuit queue shared across all available backends.
Fig.~\ref{fig:dispatch-logic} illustrates the logic of a single dispatch tick.

At each tick, the QMS broker inspects the system state and constructs an immutable snapshot capturing two categories of information: (i)~the set of pending circuits in the unified queue, ordered by arrival time, and (ii)~the runtime state of each registered backend, including the number of currently executing circuits and the latest hardware calibration data.

This snapshot is passed to the multi-device scheduling policy, which returns a device-to-circuit mapping: for each device, an ordered list of circuit identifiers to dispatch.
The contract requires that each circuit appears in at most one device's list.
In the fidelity-aware policy, the iteration over the pending circuits proceeds sequentially: for each circuit, the per-device fidelity estimates are retrieved from the GNN (either pre-computed and attached to the circuit metadata, or obtained via live inference at decision time) and the scoring function of Eq.~\eqref{eq:score} is applied against the load shares of Eq.~\eqref{eq:load_share} accumulated at the moment of each individual assignment.
Since the load counter is updated after every assignment, the load penalty adapts within the same tick: as a preferred device accumulates load share, subsequent circuits with weaker device preferences are naturally redistributed to less-loaded backends.

Once the mapping is determined, the QMS dispatches each assigned circuit to its target backend through three sequential steps: (i)~transpilation against the device's native gate set and coupling map, (ii)~submission of the transpiled circuit for execution, and (iii)~asynchronous polling for results.
Because each circuit is transpiled only for the device selected by the policy, the framework avoids the overhead of compiling every circuit for all available QPUs: the core efficiency gain over exhaustive device selection.

This architecture decouples the scheduling logic from the execution infrastructure: any policy conforming to the multi-device interface can be substituted without modifying the dispatch or execution pipeline, enabling systematic comparison of alternative strategies under identical conditions.

\section{Experimental Settings}
\label{sec:settings}
To verify the effectiveness of the proposed approach, we conduct a twofold analysis. First, we evaluate the predictive performance of the trained GNN model in terms of fidelity estimation accuracy. Second, we assess the practical applicability of the model by integrating it into a quantum circuit scheduling framework, where it is employed to guide the assignment of circuits to the most suitable target devices.

We evaluate our framework under realistic conditions by emulating the Leibniz-Rechenzentrum (LRZ) quantum computing environment, which provides access to multiple QPUs. Specifically, we consider two IQM superconducting devices: one comprising 53 qubits (EQE1) and the other 20 qubits (QExa20). For each device, the per-gate, per-qubit fidelity values are available and used to characterize the noise profile of the hardware.
We selected these two devices because their noise profiles are publicly available, which is essential for the development of this work. Despite sharing the same underlying technology, the two devices exhibit different noise profiles, resulting in diverse fidelity performance across quantum circuits. The proposed fidelity prediction model is not limited to the devices considered in this work, but can be readily extended to any HPCQC system integrating devices with different technologies and gate sets and regenerating the GNN model, as similarly demonstrated in~\cite{tudisco2025gnn}, [omitted for double blind, paper accepted].

\begin{figure}[t]
    \centering
    \begin{subfigure}{0.3\linewidth}
        \centering
        \includegraphics[width=\linewidth]{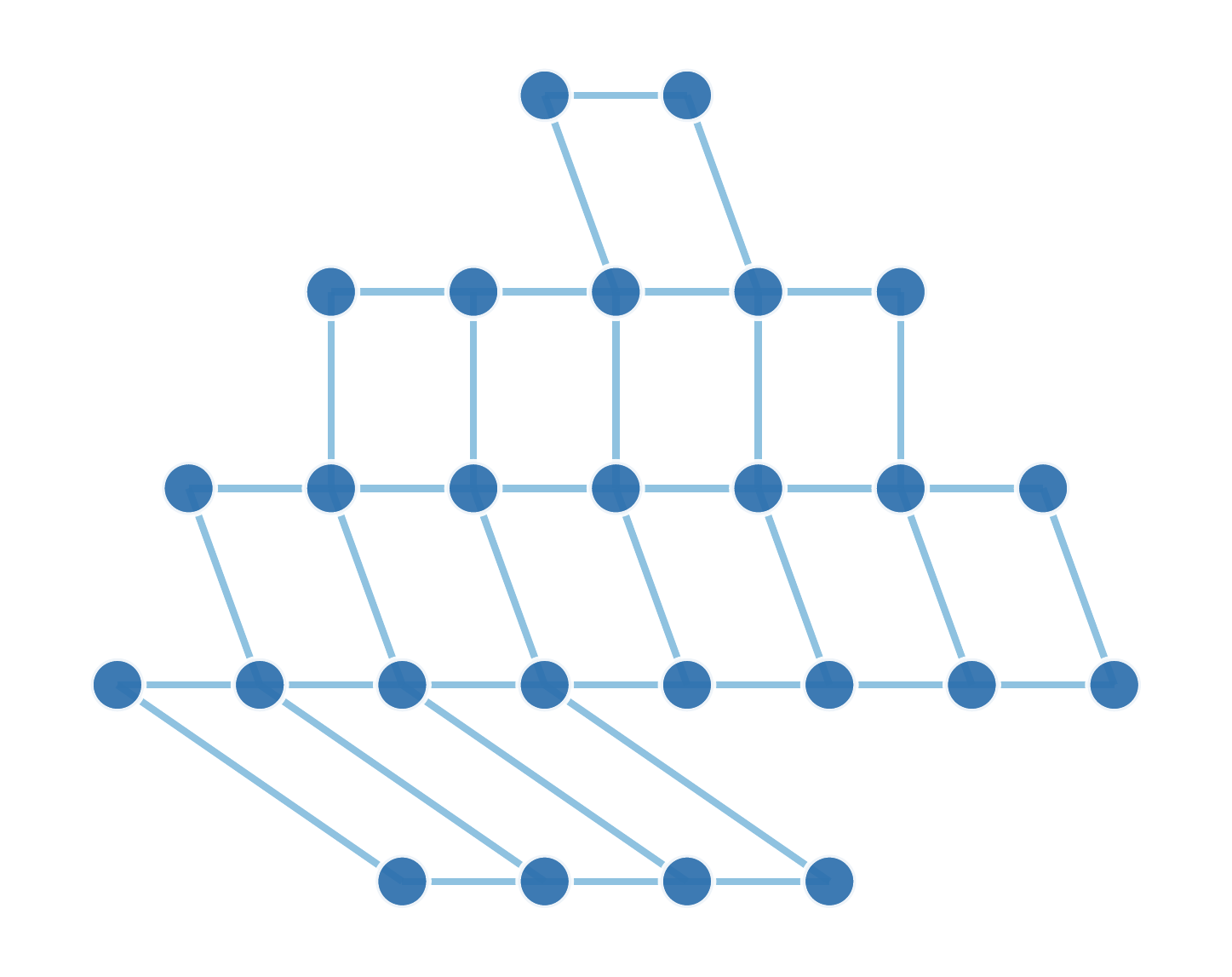}
        \caption{EQE1 Top}
    \end{subfigure}
    \hfill
    \begin{subfigure}{0.3\linewidth}
        \centering
        \includegraphics[width=\linewidth]{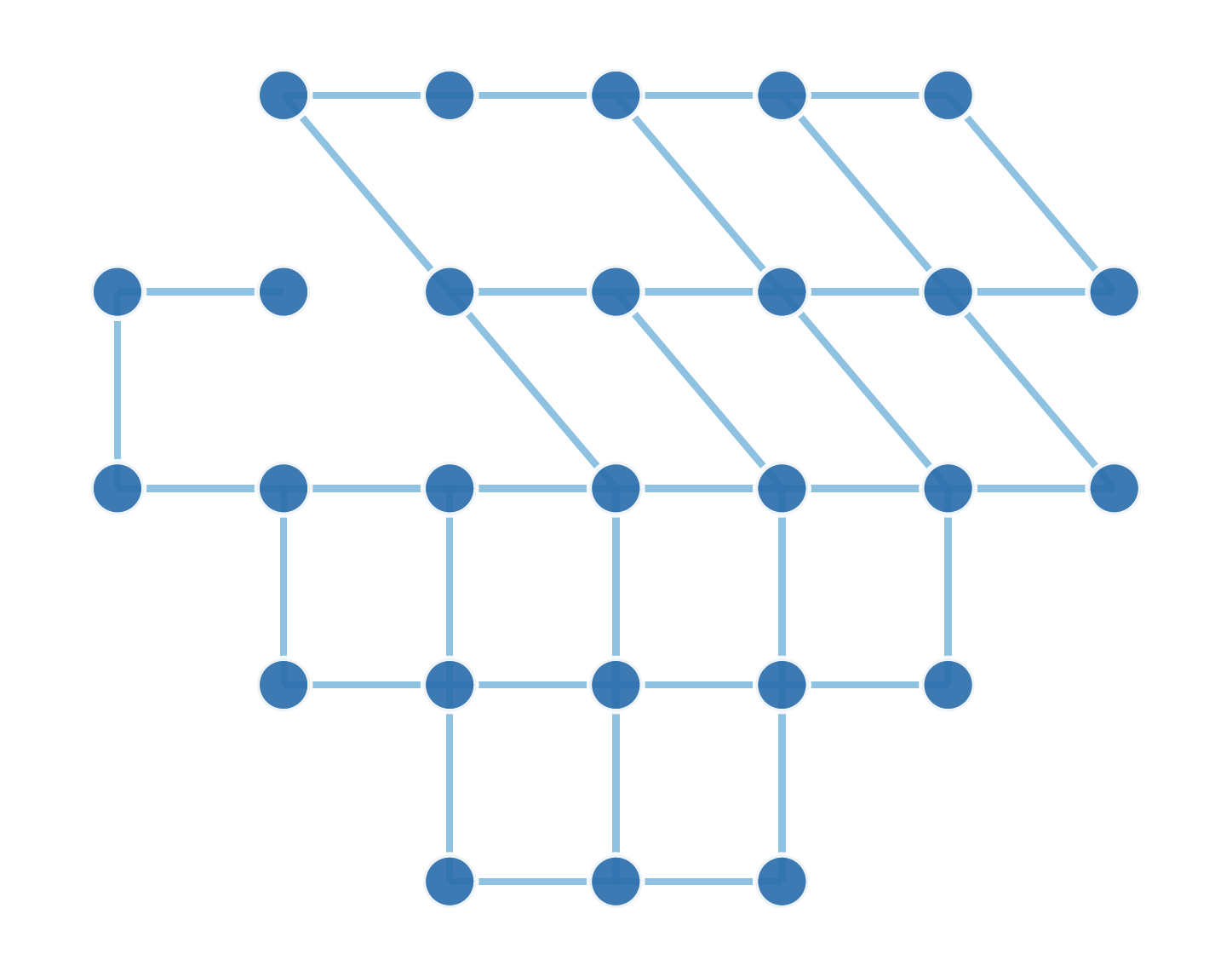}
        \caption{EQE1 Bottom}
    \end{subfigure}
    \hfill
    \begin{subfigure}{0.3\linewidth}
        \centering
        \includegraphics[width=\linewidth]{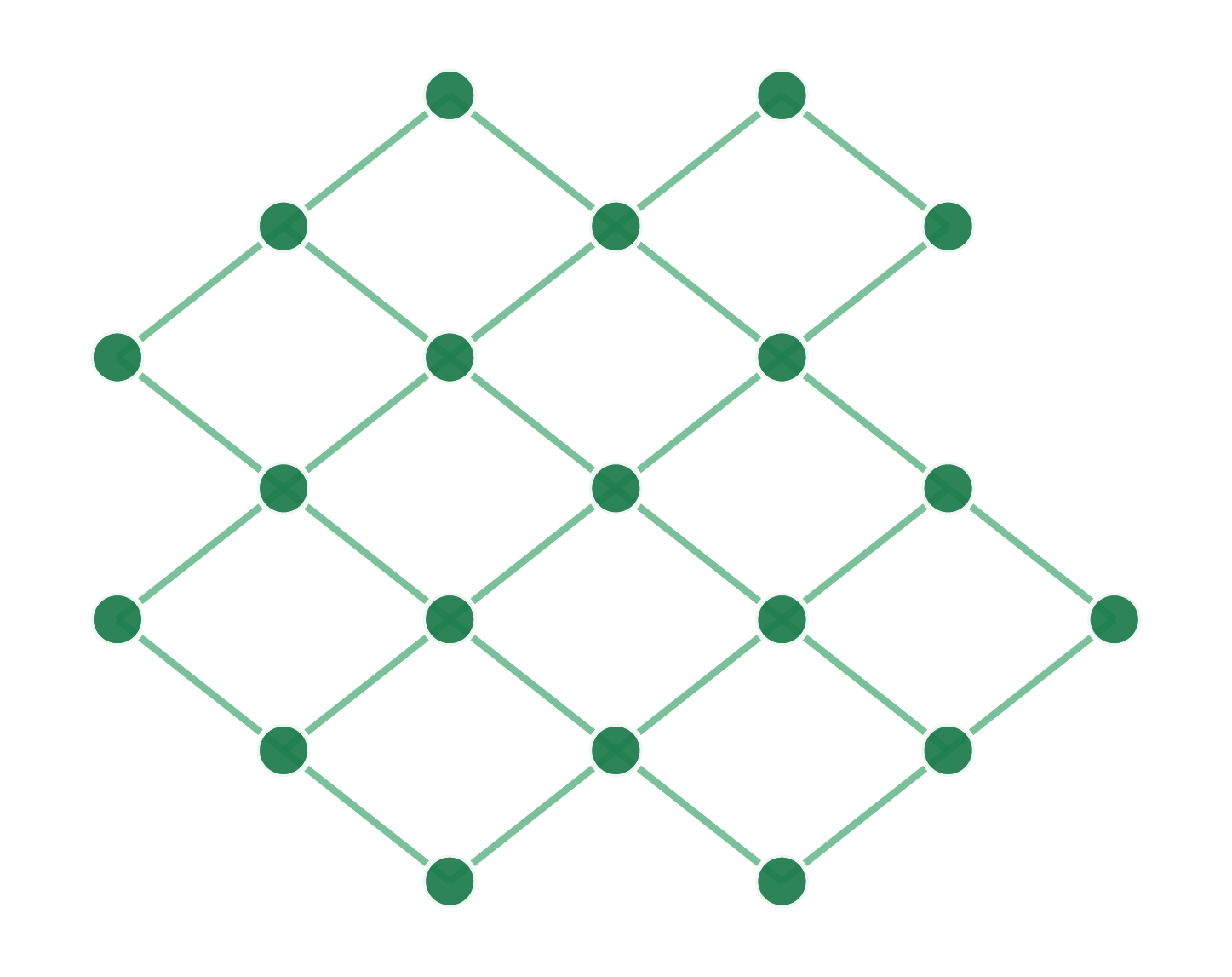}
        \caption{QExa20}
    \end{subfigure}
    \caption{Coupling maps of the target devices.}
    \label{fig:coupling-maps}
\end{figure}

\paragraph*{Device Partitioning} To further stress-test the scheduling capabilities of our framework and to better observe the effects of parallelization across a larger pool of devices, the 53-qubit device is partitioned into two independent sub-devices of 26 and 27 qubits, by performing a horizontal cut through the center of its layout. Seven cross-boundary edges are severed by the cut. Both sub-devices inherit the per-qubit calibration data from the parent device, re-indexed to contiguous qubit labels. This effectively simulates a scenario with three available QPUs and allows us to observe how the scheduler distributes circuits across devices of comparable size but different noise characteristics.
Tab.~\ref{tab:gate_fidelities} summarizes the noise characteristics of the three devices in terms of mean, minimum, and maximum gate fidelities for both single- and two-qubit operations, while Fig.~\ref{fig:coupling-maps} illustrates the respective coupling maps.

\paragraph*{Dataset}
For this work, we have adopted a dataset of 17056 circuits drawn from MQT Bench \cite{quetschlich2023mqtbench}, by augmenting the base suite with alternative state-preparation routines, ansatz topologies and oracle targets. Table~\ref{tab:circuit_categories} shows the number of circuits for each category.
For each circuit, device-specific compilation is performed on all target devices using the Qiskit compiler, setting its optimization level to 2. The ground-truth fidelity label $f_{c, d}$ for each circuit-device pair is then computed from hardware calibration data as shown in \cref{eq:fidelity}.
The dataset was partitioned into training, validation, and test sets using stratified sampling ensuring that the distribution of circuit quality is preserved across all splits.
Specifically, 70\% of the circuits form the training pool, which is further divided 80/20 into the actual training set and the validation set used during hyperparameter search.
The remaining 30\% constitute the test set, never seen during optimization.

\begin{table}[t]
\centering
\caption{Number of circuits per category.}
\label{tab:circuit_categories}
\begin{tabular}{lr@{\hspace{1.5em}}lr}
\toprule
\textbf{Category} & \textbf{\#} &
\textbf{Category} & \textbf{\#} \\
\midrule
vqe\_two\_local            & 3449 & qpeinexact                 & 19 \\
vqe\_real\_amp             & 2861 & wstate                     & 19 \\
vqe\_su2                   & 2858 & graphstate                 & 18 \\
qaoa                       & 2706 & bmw\_quark\_copula         & 10 \\
randomcircuit              & 1234 & draper\_qft\_adder         & 10 \\
qft                        & 760  & modular\_adder             & 10 \\
qnn                        & 760  & cdkm\_ripple\_carry\_adder & 9 \\
ae                         & 758  & full\_adder                & 9 \\
qftentangled               & 746  & half\_adder                & 8 \\
iqpe                       & 570  & vbe\_ripple\_carry\_adder  & 6 \\
grover                     & 62   & ghz\_dynamic               & 3 \\
hhl                        & 44   & multiplier                 & 3 \\
bmw\_quark\_cardinality    & 19   & qwalk                      & 3 \\
bv                         & 19   & rg\_qft\_multiplier        & 3 \\
dj                         & 19   & hrs\_cumulative\_multiplier& 2 \\
dynamic\_qft               & 19   & seven\_qubit\_steane\_code & 1 \\
ghz                        & 19   & shors\_nine\_qubit\_code   & 1 \\
qpeexact                   & 19   &                            &   \\
\midrule
\multicolumn{3}{l}{\textbf{Total (35 categories)}} &
\textbf{17056} \\
\bottomrule
\end{tabular}
\end{table}

\begin{table}[ht]
  \centering
  \caption{Gate fidelity statistics per device.}
  \label{tab:gate_fidelities}
  \begin{tabular}{l ccc ccc}
    \toprule
    &  \multicolumn{3}{c}{1-qubit fidelity (\%)}
    & \multicolumn{3}{c}{2-qubit fidelity (\%)} \\
    \cmidrule(lr){2-4} \cmidrule(lr){5-7}
    Device & Mean & Min & Max & Mean & Min & Max \\
    \midrule
    EQE1\_Top &  99.92 & 99.17 & 99.97 & 96.23 & 81.43 & 98.88 \\
    EQE1\_Bottom  & 99.94 & 99.59 & 99.98 & 96.78 & 84.31 & 99.13 \\
    QExa20 &  99.61 & 96.29 & 99.89 & 98.26 & 94.63 & 99.16 \\
    \bottomrule
  \end{tabular}
\end{table}

\paragraph*{Model's Hyperparameters}
The optimal model is selected through Bayesian optimization performed with Optuna~\cite{optuna} over 50 trials. Each trial consists of training a full model for up to 1000 epochs with early stopping, with the patience set to 50 epochs. The objective function minimized by Optuna is the minimum validation Mean Squared Error (MSE) reached across all epochs of the trial. The corresponding hyperparameters are reported in Tab.~\ref{tab:hyperparam}.
\section{Results} \label{sec:results}
In this section, we evaluate the effectiveness of the proposed approach along two complementary dimensions: the accuracy of our model in predicting the fidelity, and the ability of the scheduler to dispatch circuits across the available QPUs.

The code is available is publicly available at \url{https://github.com/1nnocenzo/pred-distr-tool.git}.
\subsection{Fidelity Estimation Model}
\begin{figure}
    \centering
    \includegraphics[width=\linewidth]{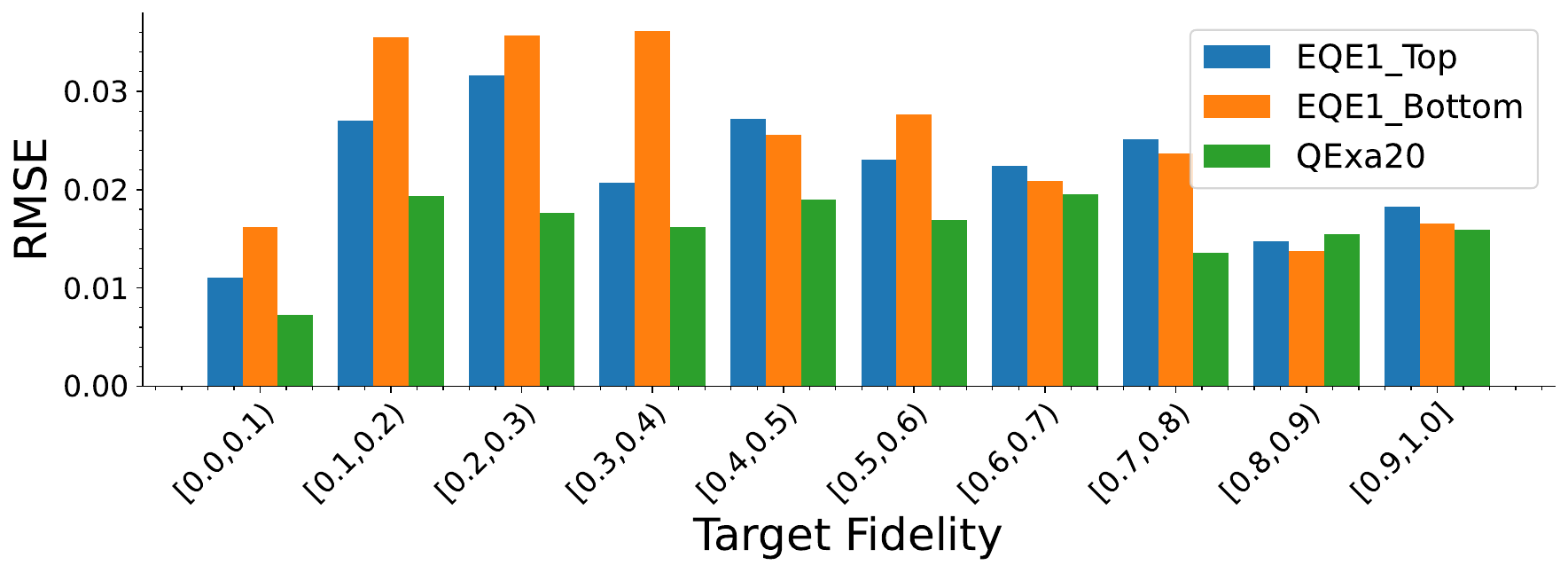}
    \caption{RMSE of the model on the test set, grouped by target fidelity.}
    \label{fig:rmse_fidelity_bin}
    \vspace{-\baselineskip}
\end{figure}

The results of the optimal model on the test set are reported in Tab.~\ref{tab:results-model-optuna}.
The results show that the model is particularly effective in predicting circuit fidelity, with low Mean Absolute Error (MAE) and Root Mean Squared Error (RMSE), on this split limited to 1.04\% and 2.00\%, respectively. 
To further characterize the model's predictive behavior, we report the RMSE grouped by target fidelity in Fig.~\ref{fig:rmse_fidelity_bin}. For each group, the RMSE remains below 4\%, with maximum errors falling for circuits whose fidelities are in the range $[0.1, 0.4]$. The model performs particularly well on the QExa20 device, where the RMSE of the fidelity prediction is about 27.5\% lower than the overall RMSE.

\begin{figure}
    \centering
    \includegraphics[width=\linewidth]{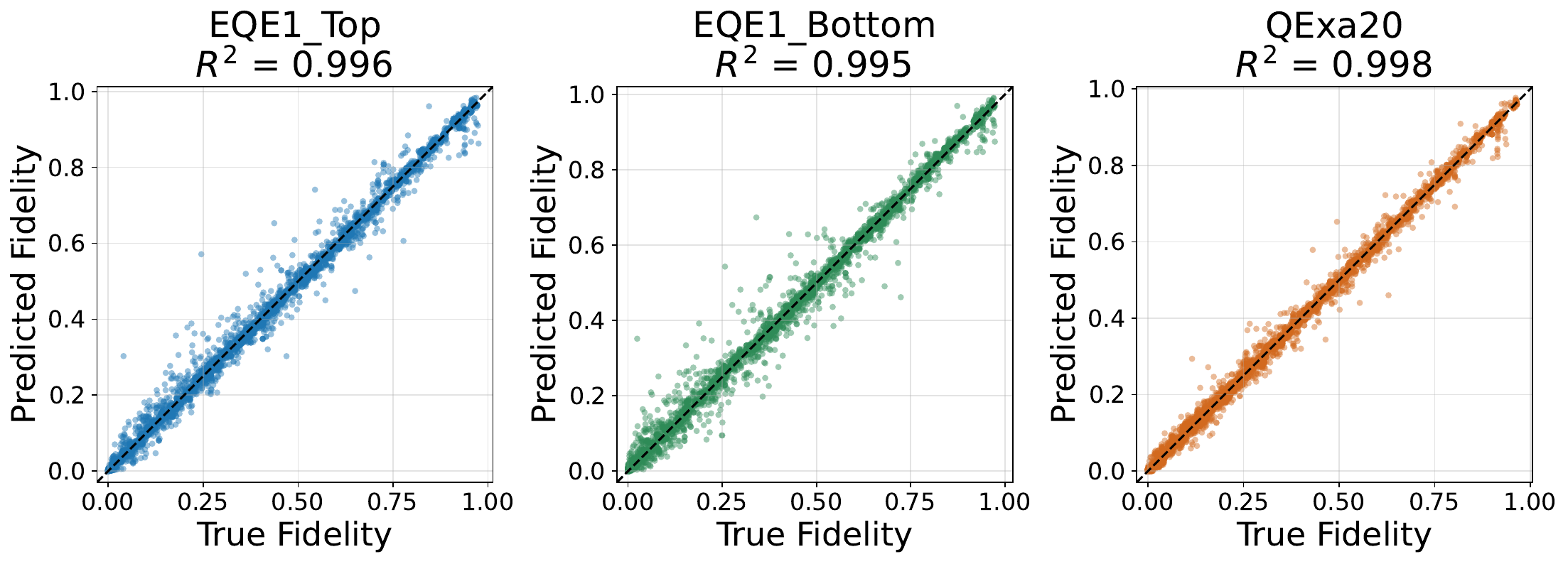}
    \caption{Predicted vs. ground-truth fidelity for each circuit–device pair on the test set.}
    \label{fig:results_prediction}
\end{figure}

To further assess the quality of the model, we compare the predicted and target fidelities on the test set for the three target devices, shown in Figure~\ref{fig:results_prediction}. Each panel reports the regression performance for a single device, with the dashed diagonal denoting perfect agreement. Across the full fidelity range, the predictions cluster tightly around the identity line, indicating that the model captures device-specific fidelity trends with little systematic bias. This confirms that the model can accurately predict circuit fidelities spanning the entire range of observed values.

\begin{table}[t]
    \centering
    \caption{Regression performance metrics of the GNN model
             on the test set, reported per device and overall.}
    \label{tab:results-model-optuna}
    \begin{tabular}{lcccc}
        \toprule
        \textbf{Device}
            & \textbf{MAE}
            & \textbf{MSE}
            & \textbf{RMSE}
            & \(\boldsymbol{R^2}\) \\
        \midrule
        EQE1\textsubscript{Top} 
            & 0.0105
            & \(4.216\times 10^{-4}\)
            & 0.0205
            & 0.9959 \\
        EQE1\textsubscript{Bot} 
            & 0.0121
            & \(5.673\times 10^{-4}\)
            & 0.0238
            & 0.9948 \\
        QExa20 
            & 0.0086
            & \(2.117\times 10^{-4}\)
            & 0.0145
            & 0.9977 \\
        \midrule
        \textbf{Overall}
            & \textbf{0.0104}
            & \(\mathbf{4.002\times 10^{-4}}\)
            & \textbf{0.0200}
            & \textbf{0.9962} \\
        \bottomrule
    \end{tabular}
\end{table}

\subsection{Scheduling}
\label{sec:meth-scheduler-results}

We evaluate the scheduling policy in a simulated multi-device setting, using the ground-truth fidelities of \cref{eq:fidelity} evaluated on the compiled test circuits and GNN-predicted fidelities obtained from the high-level circuits.
Four policies are compared:

\begin{itemize}
    \item \textbf{Oracle}: assigns each circuit to the device with the highest ground-truth fidelity, disregarding load;
    \item \textbf{GT-Weighted}: applies Eq.~\eqref{eq:score} with ground-truth fidelities, providing an upper bound for the prediction-dependent policy;
    \item \textbf{GNN}: applies the same scoring formula with GNN-predicted fidelities, while all metrics are evaluated against ground truth;
    \item \textbf{Round-Robin}: distributes circuits across devices without fidelity information.
\end{itemize}

The GNN and GT-Weighted policies are swept over an $11$-point grid $w \in \{0.0, 0.1, \ldots, 1.0\}$ on the full test split of $5117$ circuits, dispatched to the three target devices (EQE1\_Top, EQE1\_Bottom, QExa20). Oracle and Round-Robin are weight-independent by construction and serve as constant references throughout.



\subsubsection{Parallelism evaluation}

\begin{figure*}[t]
  \centering
  \includegraphics[width=\textwidth]{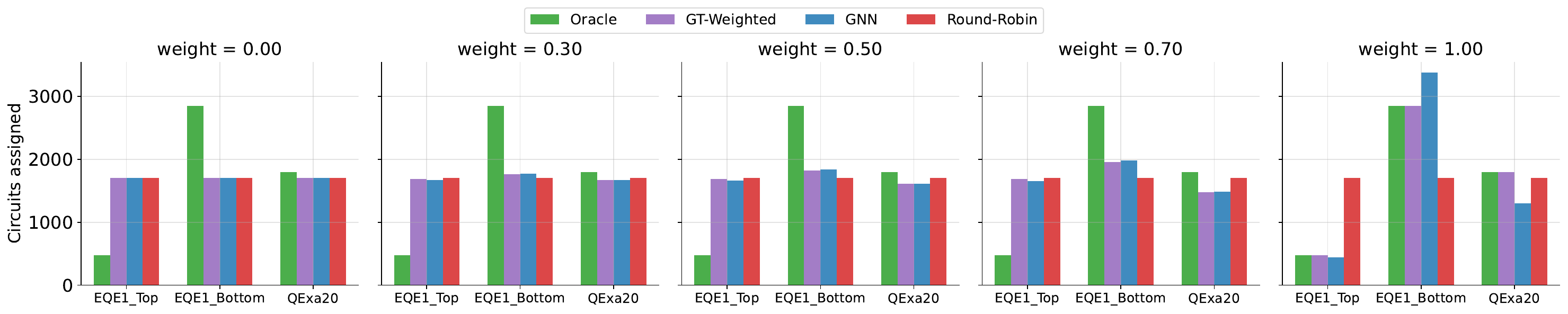}
  \caption{Load distribution across devices for each policy, shown on a reduced grid $w \in \{0.0, 0.3, 0.5, 0.7, 1.0\}$ for legibility. Round-Robin (red) is weight-independent and stays uniform across the three backends. The Oracle (green) is also weight-independent and exposes the structural bias of the dataset, concentrating 2846 circuits on EQE1\_Bottom and only 474 on EQE1\_Top. The GT-Weighted (purple) and GNN (blue) policies interpolate between these regimes as $w$ grows: at $w = 0$ the GNN coincides with Round-Robin (round-robin fallback), and the distribution skews progressively toward the Oracle profile as $w \to 1$.}
  \label{fig:load_balance}
\end{figure*}

The three-device configuration enables concurrent circuit execution on EQE1\_Top, EQE1\_Bottom, and QExa20. However, the Oracle reveals a structural bias in the dataset: 55.6\% of circuits achieve their highest ground-truth fidelity on EQE1\_Bottom, with 35.1\% on QExa20 and only 9.3\% on EQE1\_Top (load-balance coefficient of variation $\mathrm{CV} = 0.697$).
A purely fidelity-optimal policy would therefore leave EQE1\_Top substantially underutilized, reducing the effective parallelism gained by device partitioning.

Figure~\ref{fig:load_balance} visualizes how each policy distributes the $5117$ circuits across the three backends as the fidelity weight $w$ varies.
At $w = 0$, both the GNN and GT-Weighted policies reduce to a uniform distribution ($\mathrm{CV} = 0.0003$, because 5117 is not divisible by 3), fully exploiting all three backends; the GNN bars at $w = 0$ exactly match the Round-Robin reference, reflecting the round-robin fallback built into the policy.
As $w$ increases, circuits with a clear device preference migrate toward their best-fidelity backend, and the share equilibrium of Eq.~\eqref{eq:equilibrium} makes this migration gradual: the GNN reaches $\mathrm{CV} = 0.033$ at $w = 0.3$, $0.069$ at $w = 0.5$ and $0.149$ at $w = 0.7$, still distributing 1652, 1982 and 1483 circuits across the three backends at the latter operating point.
Only in the immediate neighborhood of $w = 1$, where the load term of Eq.~\eqref{eq:score} vanishes, does the allocation collapse onto the Oracle-like profile ($\mathrm{CV} = 0.884$, with 3375 circuits on EQE1\_Bottom).

In a practical setting with finite execution capacity, the Oracle's concentration, which places 2846 of $5117$ circuits on a single device while sending only 474 to EQE1\_Top, creates a load imbalance that may leave backends idle while circuits queue on the preferred device.
The GNN policy at moderate weights approaches oracle-level fidelity while distributing work far more evenly across all three devices, providing a more practical operating point for throughput-sensitive HPCQC workloads.


\subsubsection{Fidelity results}

\begin{figure}[t]
  \centering
  \includegraphics[width=\columnwidth]{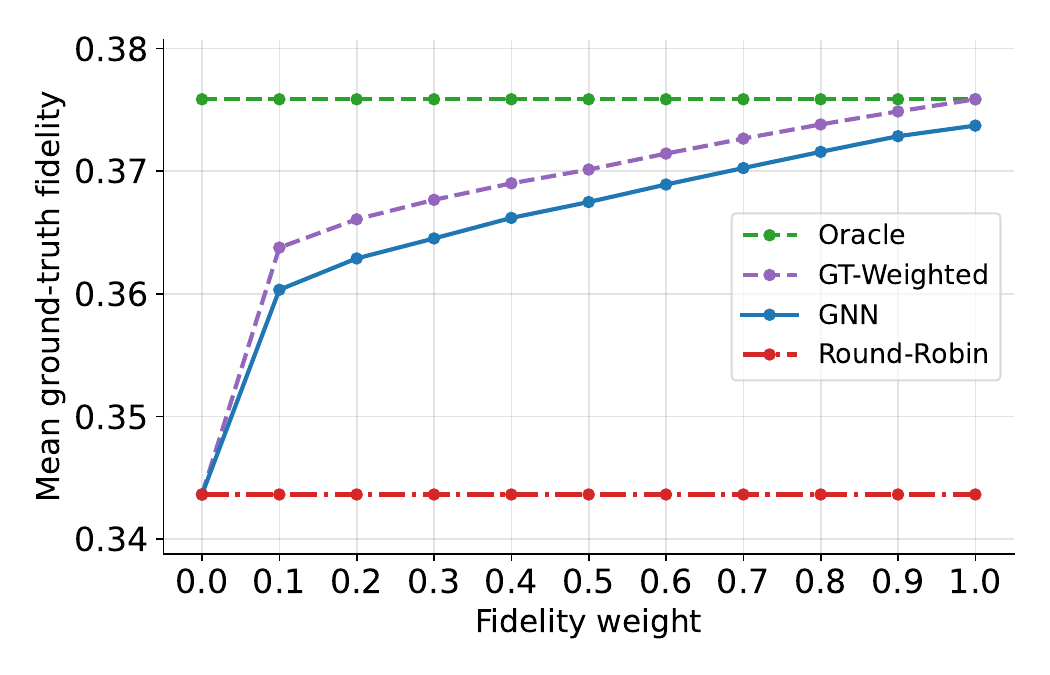}
  \caption{Overall mean ground-truth fidelity versus the fidelity weight $w$, over the $11$-point sweep. Round-Robin ($0.3436$) and the Oracle ($0.3759$) are weight-independent and bound the attainable range, a span of $0.032$ on the $[0,1]$ fidelity scale; the vertical axis is scaled to that range to resolve the policy curves, and absolute differences should be read against it. At $w = 0$ the GNN falls back to round-robin by design, so the two coincide. For $w > 0$ both prediction-aware policies increase monotonically, the GNN remaining within $0.0034$ of the GT-Weighted upper bound at every weight.}
  \label{fig:sweep}
\end{figure}

\begin{table}[t]
  \caption{Cross-policy metrics (GNN vs.\ GT-Weighted) across the fidelity weight sweep, shown on a reduced grid for compactness. $\bar{r}$: mean regret; Med.\ $r$: median regret; $r{=}0$: fraction of circuits with zero regret. The latter slightly exceeds the agreement rate because a minority of circuits attain identical ground-truth fidelity on two devices, so a disagreement in device choice costs nothing.}
  \label{tab:cross_policy}
  \centering
  \begin{tabular}{rrrrrr}
    \toprule
    $w$ & Agree (\%) & $\bar{r}$ & Med.\ $r$ & Max $r$ & $r{=}0$ (\%) \\
    \midrule
    0.00 & 100.0 & 0.0000 & 0.0000 & 0.000 & 100.0 \\
    0.30 &  68.8 & 0.0031 & 0.0000 & 0.2275 &  69.0 \\
    0.50 &  72.8 & 0.0027 & 0.0000 & 0.2232 &  73.2 \\
    0.70 &  77.5 & 0.0024 & 0.0000 & 0.2232 &  78.0 \\
    1.00 &  80.1 & 0.0021 & 0.0000 & 0.3306 &  80.1 \\
    \bottomrule
  \end{tabular}
\end{table}

Table~\ref{tab:cross_policy} and Figure~\ref{fig:sweep} summarize the fidelity results.
Round-Robin establishes a weight-independent baseline at a mean fidelity of $0.3436$, and the Oracle an upper bound of $0.3759$. Both the GT-Weighted and GNN policies improve monotonically with $w$ across the whole grid, from the Round-Robin endpoint at $w = 0$ to $0.3759$ and $0.3737$ respectively at $w = 1$. The GT-Weighted curve recovers the Oracle exactly at $w = 1$, confirming that the scoring function of Eq.~\eqref{eq:score} reproduces optimal device assignments given perfect fidelity knowledge, while the GNN tracks it within $0.0034$ at every weight.

The practically relevant quantity is the exchange rate between fidelity
and load balance along the curve. Round-robin balances the load almost
perfectly ($\mathrm{CV}=0.0003$) but only reaches a mean fidelity of
$0.3436$, while the Oracle attains $0.3759$: the entire fidelity margin
that any policy can trade load imbalance for lies in this band.
Within it, the proposed policy captures most of the available gain early.
At $w = 0.5$ it reaches $0.3675$ ($+7.0\%$ over round-robin) with
$\mathrm{CV} = 0.069$, and at $w = 0.7$ it reaches $0.3703$ ($+7.8\%$)
with $\mathrm{CV} = 0.149$. Beyond this point the marginal returns
diminish sharply: the last $0.9\%$ of mean fidelity (up to $0.3737$ at
$w = 1.0$) is bought only by letting the CV grow to $0.884$. From $w = 0.5$ onwards the policy therefore operates in the regime
where it retains nearly all of the fidelity advantage over a static
assignment while keeping all three devices in use, which is precisely
where a fidelity-aware scheduler is preferable to round-robin.


To isolate the impact of using predicted fidelities in place of ground-truth values, we compare GNN assignments directly against the GT-Weighted policy, which applies the same scoring formula with exact fidelity knowledge.
We report two metrics: the \emph{device agreement rate} (fraction of circuits assigned to the same device by both policies) and the \emph{fidelity regret} $r(c) = f_{\mathrm{GT\text{-}W}}(c) - f_{\mathrm{GNN}}(c)$, measuring per-circuit fidelity loss attributable to prediction error.
At $w = 0$, both policies reduce to round-robin, yielding 100\% agreement and zero regret by construction.
For $w > 0$ the agreement rate increases monotonically with the fidelity weight, from 68.8\% at $w = 0.3$ to 80.1\% at $w = 1$: as the fidelity term comes to dominate the score, both policies converge on the same per-circuit ranking, and the residual disagreement isolates the effect of prediction error alone.
Mean regret decreases monotonically over the same range, from $0.0031$ to $0.0021$, with a median of exactly zero throughout, confirming that the large majority of circuits land on the GT-Weighted--optimal device.
The residual ${\sim}20\%$ disagreement at $w = 1$ is concentrated on circuits whose inter-device fidelity spreads are narrow, so that a suboptimal device choice is inexpensive: the median regret is zero, the $95$th percentile is $0.011$, and only $1.15\%$ of the circuits incur a regret above $0.05$. The distribution is nonetheless long-tailed, and the maximum observed regret reaches $0.331$ on an individual circuit, so the aggregate figures should not be read as a per-circuit bound.


\subsubsection{Scheduling overhead}

\begin{figure}[t]
  \centering
  \includegraphics[width=\columnwidth]{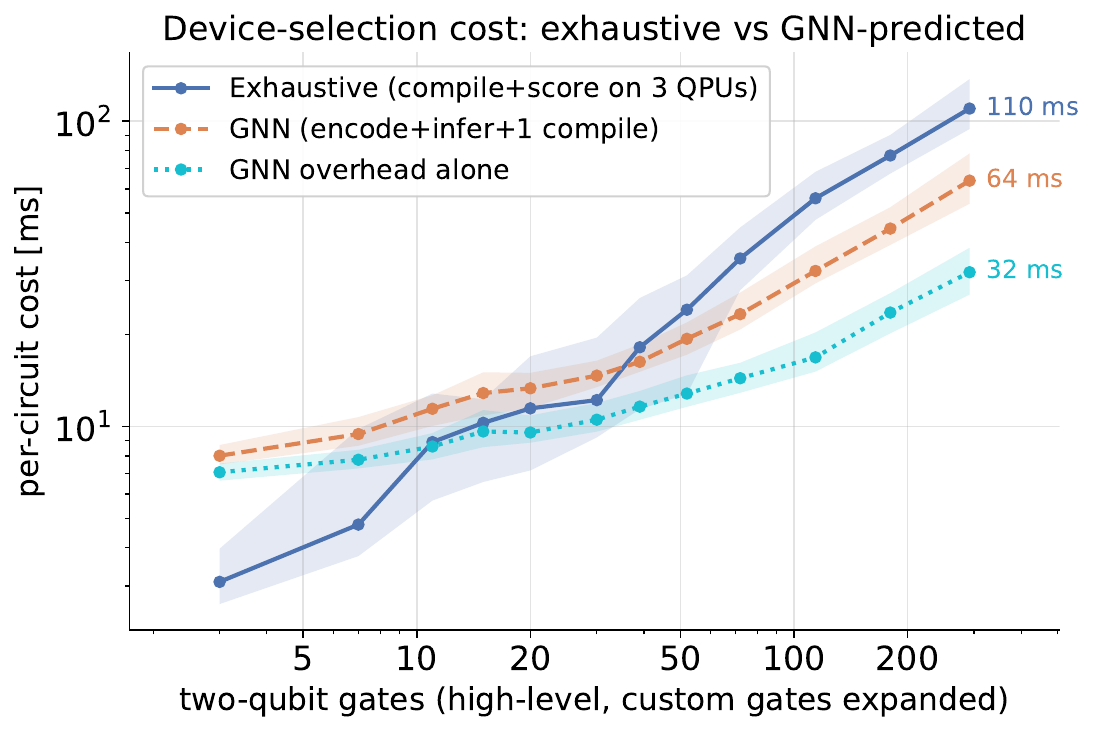}
  \caption{Median per-circuit wall-clock cost of the two device-selection strategies against the number of two-qubit gates in the submitted circuit, with inter-quartile bands, on logarithmic axes; circuits are grouped into twelve equal-count bins. Both paths grow with the entangling-gate count, but the exhaustive cost grows faster because it pays the layout and routing effort on all $D$ devices, and the two cross at about $35$ two-qubit gates. The encoding and inference overhead (dotted) is the floor of the prediction-based path. The entangling-gate count is used rather than circuit width, total gate count, or depth because it is the quantity that actually drives layout and routing cost: on this split its rank correlation with the exhaustive cost is $0.87$, against $0.55$ for width, $0.55$ for depth and $0.52$ for total gate count, and it is the only one of the four under which the binned cost is monotone. Medians are taken within bins, so the separation shown is wider than the aggregate speedup, which is diluted by the overlap of the two distributions.}
  \label{fig:timing}
\end{figure}

The central motivation of this work is avoiding the $k \times D$ compilations required by exhaustive device selection, so we measure both paths directly. Figure~\ref{fig:timing} reports the per-circuit wall-clock breakdown, as a function of the two-qubit gate count. Timings were collected on the same test split, excluding QASM parsing from both sides and constructing the pass managers and the model once outside the measurement loop, so that only the per-circuit cost is compared. Of the $5117$ test circuits, all are timed except for the Grover instances, which are omitted because their transpiled size exceeds the others by four orders of magnitude and would dominate every aggregate.

Over the whole split, the prediction-based path is faster both in the median and mean with respect to exhaustive compilation. The two additional costs of the fast path, DAG encoding and inference, together amount to $11.57$\,ms in the median, comparable at $D = 3$ to the two compilations it avoids. They are not constant, growing from $6.94$\,ms at two qubits to $18.27$\,ms at twenty, but they grow more slowly than the exhaustive cost, and it is this difference in slope that produces a crossover. As Figure~\ref{fig:timing} shows, the two paths cross at about $35$ two-qubit gates: below that the exhaustive route is cheaper, while in the densest bin, at a median of $292$ two-qubit gates, exhaustive selection costs $110$\,ms against $64$\,ms for the prediction-based path, a $1.72\times$ saving. The same trend holds in the other size measures, with the advantage reaching $1.73\times$ at $20$ qubits and $1.65\times$ in the deepest depth quartile. Conversely, on the shallowest quartile, the exhaustive path remains cheaper, and $47\%$ of the circuits in the split are individually faster to compile exhaustively than to predict. The gain scales with $D$, since the exhaustive cost is linear in the number of devices while the prediction cost is not, so the three-device configuration studied here represents a lower bound on the achievable saving.

Training-label generation is itself an exhaustive process and must be accounted for. Compiling and scoring on all three devices the $\sim 12$ thousand timed circuits that fall outside the test split costs $904.5$\,s in total. Against a mean saving of $9.96$\,ms per dispatched circuit, this one-off cost is amortized after approximately $9.1 \times 10^{4}$ dispatched circuits, beyond which the framework yields a net saving over exhaustive selection at $D = 3$; the break-even point falls as $D$ grows.

Two caveats bound these figures. Inference is measured at batch size one on CPU, matching a per-circuit dispatch decision but not the batched regime in which a meta-scheduler would realistically operate; and the encoding stage is a pure-Python DAG traversal. Both are implementation rather than algorithmic costs, so the reported speedup should be read as a conservative estimate.

\section{Conclusions}
\label{sec:conclusions}
This work presents a fidelity-aware scheduling framework for heterogeneous multi-QPU systems. At its core, a GNN model exploits the DAG representation of quantum circuits to accurately predict the expected fidelity of each circuit on each target device, prior to compilation. The resulting per-device fidelity estimates are then fed to a tunable scheduler, which assigns each circuit to a target QPU according to a configurable policy that allows the user to control the trade-off between execution fidelity and workload parallelism across the available devices. The experimental evaluation demonstrates the effectiveness of the proposed approach: the framework achieves a mean batch fidelity improvement over a round-robin baseline, while avoiding the compilation overhead associated with exhaustive multi-device selection.

As future work, a natural next step is to validate the proposed framework on real multi-QPU infrastructures with concrete quantum circuit batches to study its behavior on the application level. 
Another direction is to evaluate the model across a broader set of quantum technologies and native gate sets, assessing how the learned circuit representations generalize across substantially different hardware platforms.


\section*{Acknowledgment}
The work was funded by the Munich Quantum Valley (MQV), which is supported by the Bavarian State Government with funds from the Hightech Agenda Bayern.
Furthermore, this research is funded by BMIMI, BMWET, the state of Upper Austria and the State of Tyrol within the COMET module Quantum Algorithm Engineering (FFG Grant no. 923923) managed by Austrian Research Promotion Agency FFG.
Furthermore, the authors would like to thank Hossam Ahmed of the Leibniz Supercomputing Center (LRZ) for his invaluable support in providing and interpreting the device data used in this research.
The authors acknowledge HPC@PoliTo for providing computational resources.
Large Language Models have been used to support the development of the scripts employed in the experimental evaluation and to rephrase parts of the manuscript text. All content has been reviewed and validated by the authors, who take full responsibility for it.
\bibliographystyle{IEEEtran}
\bibliography{bibliography}

\end{document}